\documentclass[aps,prc,10pt,twocolumn,superscriptaddress,showpacs]{revtex4-1}
\usepackage{graphicx}
\usepackage{color}
\usepackage{bm}
\usepackage{amsmath}
\usepackage{multirow}

\begin{document}

\title{Representation of compounds for machine-learning prediction of physical properties}
\author{Atsuto \surname{Seko}}
\email{seko@cms.mtl.kyoto-u.ac.jp}
\affiliation{Department of Materials Science and Engineering, Kyoto University, Kyoto 606-8501, Japan}
\affiliation{Center for Elements Strategy Initiative for Structure Materials (ESISM), Kyoto University, Kyoto 606-8501, Japan}
\affiliation{JST, PRESTO, Kawaguchi 332-0012, Japan}
\affiliation{Center for Materials Research by Information Integration, National Institute for Materials Science, Tsukuba 305-0047, Japan}
\author{Hiroyuki \surname{Hayashi}}
\affiliation{Department of Materials Science and Engineering, Kyoto University, Kyoto 606-8501, Japan}
\affiliation{Center for Materials Research by Information Integration, National Institute for Materials Science, Tsukuba 305-0047, Japan}
\author{Keita \surname{Nakayama}}
\affiliation{Department of Materials Science and Engineering, Kyoto University, Kyoto 606-8501, Japan}
\author{Akira \surname{Takahashi}}
\affiliation{Department of Materials Science and Engineering, Kyoto University, Kyoto 606-8501, Japan}
\author{Isao \surname{Tanaka}}
\affiliation{Department of Materials Science and Engineering, Kyoto University, Kyoto 606-8501, Japan}
\affiliation{Center for Elements Strategy Initiative for Structure Materials (ESISM), Kyoto University, Kyoto 606-8501, Japan}
\affiliation{Center for Materials Research by Information Integration, National Institute for Materials Science, Tsukuba 305-0047, Japan}
\affiliation{Nanostructures Research Laboratory, Japan Fine Ceramics Center, Nagoya 456-8587, Japan}

\date{\today}

\begin{abstract}
The representations of a compound, called ``descriptors" or ``features", play an essential role in constructing a machine-learning model of its physical properties.
In this study, we adopt a procedure for generating a systematic set of descriptors from simple elemental and structural representations.
First it is applied to a large dataset composed of the cohesive energy for about 18000 compounds computed by density functional theory (DFT) calculation.
As a result, we obtain a kernel ridge prediction model with a prediction error of 0.041 eV/atom, which is close to the ``chemical accuracy" of 1 kcal/mol (0.043 eV/atom).
The procedure is also applied to two smaller datasets, i.e., a dataset of the lattice thermal conductivity (LTC) for 110 compounds computed by DFT calculation and a dataset of the experimental melting temperature for 248 compounds.
We examine the performance of the descriptor sets on the efficiency of Bayesian optimization in addition to the accuracy of the kernel ridge regression models.
They exhibit good predictive performances.
\end{abstract}

\maketitle
\section{Introduction}
A data-driven machine learning approach is expected to be used to make prediction models of target physical properties of interest and classification models of target classes of properties.
Therefore, machine-learning techniques have been increasingly used for the exploration of materials and structures from a huge number of candidates\cite{Curtarolo_PRL_2003,Fischer_NM_2006,hautier2010finding,hautier2010data,PhysRevLett.115.205901,doi:10.1146/annurev.matsci.38.060407.130217,carrete2014finding,PhysRevLett.117.135502,pilania2013accelerating,PhysRevB.89.094104,pilania2016machine,PhysRevB.93.085142,Kiyoharae1600746,PhysRevB.93.054112,ward2016general,kim2016organized,isayev2015materials}
and/or for extracting meaningful information and patterns from existing data such as machine-learning interatomic potentials and so forth\cite{Lorenz2004210,behler2007generalized,bartok2010gaussian,PhysRevB.90.024101,PhysRevLett.114.105503,casp:seko,seko2014machine,balachandran2015materials,snyder2012finding,mueller2014origins,caccin2015framework,PhysRevB.93.115104,nelson2013compressive,PhysRevLett.113.185501,PhysRevB.92.054301}.
A key factor in controlling the performance of a machine-learning-based approach is how compounds are represented in a dataset. 
Representations of compounds are called ``descriptors" or ``features". 

Candidate compound descriptors are quantities obtained from first-principles physical properties such as volume, cohesive energy, elastic constants, dielectric constants and so forth.
Although a few first-principles databases are available, the numbers of compounds and physical properties in the databases are still limited.
Nevertheless, if one can discover a set of descriptors that can explain a target property well, a robust prediction model of a target property can be derived.
In addition to first-principles properties, one can use quantities derived from simple representations of elements and structures of compounds as descriptors.

Many candidate elemental representations can be found in the literature.
These are
intrinsic quantities such as atomic number and ionization energy,
heuristic quantities such as electronegativity\cite{doi:10.1021/ja01348a011} and ionic radius\cite{shannon1976revised}
and physical properties of elemental substances such as melting and boiling points.
Such elemental representations have already been used in many studies on the machine learning prediction.
Other candidates are the chemical composition or a binary digit representing the presence of each element in a compound\cite{PhysRevLett.115.205901}.
Also, an elemental or ionic similarity defined by crystal structure database entries has been proposed\cite{PhysRevB.88.224107}.
Many structural representations that are not generally intended for application to machine learning have also been proposed in the literature.
They include the simple coordination number, Voronoi polyhedron of a central atom, angular distribution function (ADF), bond-orientational order parameter (BOP)\cite{PhysRevB.28.784} and radial distribution function (RDF).
Some of them and their extended forms have already been applied to machine-learning predictions\cite{PhysRevB.89.205118,von2013representation,PhysRevB.90.054102}.
Candidate structural representations have also been proposed in the context of machine learning interatomic potential\cite{bartok2013representing}.

Moreover, in the usual situation that a dataset covers a wide range of chemical compositions and crystal structures, it is natural to consider a combined form of elemental and structural representations as a descriptor.
Such descriptors have not been proposed except for the Coulomb matrix\cite{PhysRevLett.108.058301}, its extended forms\cite{faber2015crystal}, the X-ray diffraction pattern
and some functions based on multiple elemental representations\cite{PhysRevLett.114.105503,ghiringhelli2016learning}, because it is not easy to find a good descriptor derived from elemental and structural representations.
Therefore, a systematic procedure for generating a set of compound descriptors from simple representations is strongly required.

In this study, we demonstrate an approach to derive a set of descriptors for a compound from atomic representations, which can be applied not only to crystalline systems but also to molecular systems.
This approach enables us to generate a systematic set of descriptors composed of elemental and structural representations satisfying the following features.
(1) Compounds with a wide range of chemical compositions are expressed by same-dimensional descriptors.
(2) Compounds with a wide range of crystal structures are expressed by same-dimensional descriptors. 
This is an important feature because unit cells of different crystals are not composed of the same number of atoms.
(3) A set of descriptors satisfies translational and rotational invariance and other invariances required for all compounds included in the dataset.

We apply this approach to a large dataset and two small datasets of physical properties. 
The large dataset is composed of the cohesive energy for about 18000 compounds computed by density functional theory (DFT) calculation\cite{PhysRev.136.B864,PhysRev.140.A1133}.
The two small datasets correspond to a dataset of the lattice thermal conductivity (LTC) for 110 compounds computed by DFT calculation and a dataset of the experimental melting temperature for 248 compounds.
Using the datasets, we examine the performance of descriptors in terms of the prediction error for test data of the kernel ridge models and the performance of the Bayesian optimization.

This paper is organized as follows.
Section \ref{info_descriptors:sec_dataset} gives the detail of the datasets used in this study. 
Section \ref{info_descriptors:sec_regmethods} describes the regression methods and the procedure for Bayesian optimization adopted in this study. 
Section \ref{info_descriptors:sec_descriptors} shows how to derive compound descriptors from simple elemental and structural representations.
Section \ref{info_descriptors:sec_results} gives results for the performance of the descriptors for kernel ridge models and Bayesian optimization.
Finally, we conclude in Sec. \ref{info_descriptors:sec_conclusion}.

\section{Datasets}
\label{info_descriptors:sec_dataset}

\subsection{DFT cohesive energy (18093 compounds)}

\begingroup
\squeezetable
\begin{table*}[tbp]
\caption{
Elements and their valences included in the DFT dataset of cohesive energy.
Prototype structures of compounds included in the DFT dataset are also shown.
We adopt prototype structures for which many entries are registered in the ICSD.
}
\label{info_descriptors:prototype}
\begin{ruledtabular}
\begin{tabular}{llc}
Valence & Element &\\
\hline
1+ & Li, Na, K, Rb, Cs &\\
2+ & Be, Mg, Ca, Sr, Ba, Zn, Cd, Hg &\\
3+ & Al, Ga, In, Sc, Y, La &\\
1$-$ & F, Cl, Br, I &\\
2$-$ & O, S, Se, Te &\\
3$-$ & N, P, As, Sb &\\
\\
Formula & Prototype structure & Number of compounds\\
\hline
AX & NaCl, ZnS, ZnO, NiAs, MnP, FeB, CsCl, TlI & 608 \\
AX$_2$ & CaF$_2$, CaCl$_2$, $\alpha$-PbO$_2$, CdI$_2$, La$_2$Sb, Rutile-TiO$_2$, PbCl$_2$, Pyrite-FeS$_2$, Marcasite-FeS$_2$ & 468 \\
AX$_3$ & Cementite-Fe$_3$C, YF$_3$, Na$_3$As, ReO$_3$, BiI$_3$ & 220 \\
A$_2$X$_3$ & Bi$_2$Te$_3$, La$_2$O$_3$, Sn$_2$S$_3$, Sb$_2$S$_3$, Bixbyite-Mn$_2$O$_3$ & 279 \\
ABX & Cu$_2$Sb, RbAuS, Fe$_2$P & 2214 \\
ABX$_2$ & $\alpha$-LiFeO$_2$, NaCrS$_2$, CuFeS$_2$, CuLaS$_2$, AgFeO$_2$ & 6497 \\
ABX$_3$ & Perovskite-GdFeO$_3$, Perovskite-CaTiO$_3$, BaNiO$_3$ & 2243 \\
ABX$_4$ & ZrSiO$_4$, CaWO$_4$, BiWO$_4$ & 2207 \\
AB$_2$X$_4$ & Olivine-Mg$_2$SiO$_4$, Spinel-MgAl$_2$O$_4$, K$_2$MgF$_4$, CaFe$_2$O$_4$, CrNb$_2$Se$_4$, Al$_2$CdS$_4$ & 3357 \\
\end{tabular}
\end{ruledtabular}
\end{table*}
\endgroup

The first dataset contains the cohesive energy for binary and ternary compounds computed by DFT calculation.
The compounds in the dataset correspond to exhaustive arrangements of given chemical compositions and crystal structure prototypes.
Therefore, the cohesive energy depends on both the elements and the crystal structure of the compound.
The chemical compositions are generated by considering all combinations of cations and anions listed in Table \ref{info_descriptors:prototype}, satisfying the charge neutrality condition.
Binary compounds have the compositions of AX, AX$_2$, AX$_3$ and A$_2$X$_3$, and ternary compounds have the compositions of ABX, ABX$_2$, ABX$_3$, ABX$_4$ and AB$_2$X$_4$, where elements A and B are cations and element X is an anion.
For each chemical composition, we consider several crystal structure prototypes, shown in Table \ref{info_descriptors:prototype}, included in the inorganic crystal structure database (ICSD)\cite{bergerhoff1987crystal}.
The total number of compounds is 18093, 1575 binary and 16518 ternary compounds.

We use a definition of the cohesive energy for a binary or ternary compound, normalized by the total number of atoms, expressed as
\begin{equation}
E_{\rm coh} = \frac{\left(n_{\rm A} E_{\rm A}^{\rm atom} + n_{\rm B} E_{\rm B}^{\rm atom} + n_{\rm X} E_{\rm X}^{\rm atom} \right) - E^{\rm bulk}}{n_{\rm A} + n_{\rm B} + n_{\rm X}}, 
\end{equation}
where $n_{\rm A}$, $n_{\rm B}$ and $n_{\rm X}$ denote the numbers of atoms A, B and X included in a simulation cell for the compound, respectively.
$E^{\rm bulk}$ is the total energy of the compound at the equilibrium volume.
$E_{\rm A}^{\rm atom}$, $E_{\rm B}^{\rm atom}$ and $E_{\rm X}^{\rm atom}$ are the energies of isolated atoms A, B and X, respectively.

For all 18093 compounds, DFT calculations were performed using the plane-wave basis projector augmented wave (PAW) method\cite{PAW1,PAW2} within the Perdew--Burke--Ernzerhof exchange-correlation functional\cite{GGA:PBE96} as implemented in the \textsc{vasp} code\cite{VASP1,VASP2}.
The cutoff energy was set to 400 eV.
The total energy converged to less than 10$^{-3}$ meV.
The atomic positions and lattice constants were optimized until the residual forces became less than $10^{-2}$ eV/\AA. 
To evaluate the energy of an isolated atom, a spin-polarized calculation was performed with a large periodic cell of 15 \AA\ $\times$ 16 \AA\ $\times$ 17 \AA.

\subsection{DFT lattice thermal conductivity (110 compounds)}
One of the small datasets is composed of the LTC for 110 compounds computed by DFT calculation, generated by incorporating the datasets used in Refs. \onlinecite{PhysRevLett.115.205901} and \onlinecite{togo2015distributions}.
We employed the supercell and finite-displacement approaches to obtain second-order and third-order force constants.
LTCs were calculated from phonon lifetimes, group velocities and mode-heat capacities solving the phonon Boltzmann transport equation within the relaxation time approximation.
The \textsc{phonopy}\cite{togo2015first} and \textsc{phono3py}\cite{togo2015distributions} codes were used for these phonon calculations.
Details of the theoretical background and computational procedure can be found in Refs. \onlinecite{PhysRevLett.115.205901} and \onlinecite{togo2015distributions}. 

\subsection{Experimental melting temperature (248 compounds)}
The other small dataset is composed of the experimental melting temperature for 248 binary compounds taken from Ref. \onlinecite{CRC-Handbook-Chemistry-Physics}. 
This dataset is exactly the same as that used in Ref. \onlinecite{seko2014machine}. 
The melting temperatures of the compounds in the dataset range from room temperature to 3273 K. 
In addition, transition-metal compounds are not included in the dataset to avoid complexity in the DFT calculation.
The compounds and their melting temperatures can be found in the Appendix of Ref. \onlinecite{seko2014machine}.

Since the database of melting temperatures does not contain information of the crystal structure, we estimate the stable crystal structure for each compound by DFT calculation.
Candidates crystal structures are taken from the ICSD.
When the ICSD database has a unique crystal structure for a compound, the DFT calculation is carried out for the unique crystal structure.
When the ICSD database contains multiple crystal structures for the compound, the crystal structure with the lowest energy among the structures is adopted.

\section{Regression methods}
\label{info_descriptors:sec_regmethods}

\subsection{Kernel ridge regression}
A way of measuring the performance of descriptors is to estimate the prediction error of regression models.
Kernel ridge regression (KRR)\cite{murphy2012machine} is employed in this study.
In the formalism of KRR, the observation property $y$ of point $\bm{d}$ is expressed by a kernel function for point $\bm{d}$ and training data point $\bm{d}_i$ as 
\begin{equation}
y\left(\bm{d} \right) = \sum_{i=1}^{N} \alpha_i k\left(\bm{d}, \bm{d}_i \right),
\end{equation}
where $N$ and $\alpha_i$ denote the number of training data and the contribution of training data $i$ to the prediction of the observation property, respectively.
$k\left(\bm{d}, \bm{d}_i \right)$ is a kernel function used to measure the similarity between point $\bm{d}$ and training data point $\bm{d}_i$.
Here we introduce a radial basis function (RBF) kernel, given by
\begin{equation}
k\left(\bm{d}_i, \bm{d}_j \right) = \exp \left( - \frac{|\bm{d}_i - \bm{d}_j|^2}{2 \sigma^2} \right)
\end{equation}
with a length scale of $\sigma$.
Coefficients $\bm{\alpha} = [\alpha_1, \alpha_2, \cdots, \alpha_N ]^\top$ are determined from the training data by simple matrix operations as
\begin{equation}
\bm{\alpha} = \left( \bm{K} + \lambda \bm{I}_N \right)^{-1} \bm{y},
\end{equation}
where $\lambda$ and $\bm{I}_N$ denote a regularization parameter and the $N$-dimensional identity matrix, respectively.
$\bm{y}$ denotes an $N$-dimensional vector describing the observation property of the training data.
$\bm{K}$ is a symmetric kernel matrix composed of kernel functions for all pair arrangements of the training data, expressed as
\begin{equation}
\bm{K} = 
\begin{pmatrix}
k\left(\bm{d}_1, \bm{d}_1 \right) & k\left(\bm{d}_1, \bm{d}_2 \right) & \cdots & k\left(\bm{d}_1, \bm{d}_N \right) \\
k\left(\bm{d}_2, \bm{d}_1 \right) & k\left(\bm{d}_2, \bm{d}_2 \right) & \cdots & k\left(\bm{d}_2, \bm{d}_N \right) \\
\vdots & \vdots & \ddots & \vdots \\
k\left(\bm{d}_N, \bm{d}_1 \right) & k\left(\bm{d}_N, \bm{d}_2 \right) & \cdots & k\left(\bm{d}_N, \bm{d}_N \right) \\
\end{pmatrix}
.
\end{equation}
Finally, the prediction for point $\bm{d}_*$ is derived as
\begin{equation}
y\left(\bm{d}_* \right) = {\bm k}_*^\top \left( \bm{K} + \lambda \bm{I}_n \right)^{-1} \bm{y},
\label{info_descriptors:krr_eqn}
\end{equation}
where ${\bm k}_* = \left[ k(\bm{d}_*, \bm{d}_1), \cdots, k(\bm{d}_*, \bm{d}_N) \right] ^\top$ is the vector of kernel functions for point $\bm{d}_*$ and the training examples.
The prediction model depends on the values of $\sigma$ and $\lambda$, hence we determine them by a grid-search optimization.

\subsection{Gaussian process regression}
Another way of measuring the performance of descriptors is to examine the efficiency in finding the compound showing the best observation property among the existing data whose observations are known.
We employ Bayesian optimization based on a Gaussian process (GP)\cite{Rasmussen_2006}, specified by its mean function and covariance function.
We adopt an RBF covariance for noise-free observation, given by
\begin{equation}
k\left(\bm{d}_i, \bm{d}_j \right) = \sigma_f^2  \exp \left( - \frac{|\bm{d}_i - \bm{d}_j|^2}{2 l^2} \right), 
\end{equation}
where $l$ and $\sigma_f^2$ are tuning parameters controlling the length scales for $\bm{d}$ and the observation, respectively.
The mean function $\mu$ at point $\bm{d}_*$ and the variance function $\sigma_*^2$ are given as
\begin{equation}
\mu(\bm{d}_*) = {\bm k}_*^\top \bm{K}^{-1} \bm{y}
\end{equation}
and
\begin{equation}
\sigma_*^2 = k(\bm{d}_*,\bm{d}_*) - {\bm k}_*^\top \bm{K}^{-1} {\bm k}_*,
\end{equation}
respectively. 
The mean function is exactly the same as the KRR prediction in Eqn. (\ref{info_descriptors:krr_eqn}) without the regularization term.

\subsection{Bayesian optimization}
Our procedure for Bayesian optimization is as follows.
First, a GP model is developed from two randomly selected observations taken from all the data.
The model is iteratively updated by repeatedly (i) sampling the point at which the observation property is expected to be the best and (ii) updating the model including the observation at the sampled point. 
These steps are repeated until all the data are sampled.

We have two main options when sampling a new point\cite{jones01}: we can consider the probability of improvement (PI) and the expected improvement (EI).
For a minimization problem, the former involves sampling the point at which the probability that the observation is lower than $y_{\rm best}$ is maximized, where $y_{\rm best}$ denotes the best observation among the observed data.
Therefore, sampling point $i'$ is selected by maximizing the probability formulated as
\begin{equation}
i': = \mathop{\rm argmax}\limits_{\bm{d}_*} \Phi \left(\frac{y_{\rm best}-\mu(\bm{d}_*)}{\sigma_*} \right),
\end{equation}
where $\Phi \left(y-\mu(\bm{d}_*)/\sigma_* \right)$ denotes the cumulative distribution function of $N(\mu, \sigma^2)$.
In a similar manner, the EI is formulated as
\begin{equation}
i': = \mathop{\rm argmax}\limits_{\bm{d}_*} \int_{-\infty}^{y_{\rm best}} (y_{\rm best} - y) \phi \left(\frac{y -\mu(\bm{d}_*)}{\sigma_*} \right) dy,
\end{equation}
where $\phi \left(y-\mu(\bm{d}_*) / \sigma_* \right)$ denotes the probability density function of $N(\mu, \sigma^2)$.
We apply the two options for Bayesian optimization to the LTC and melting temperature datasets.

\section{Descriptors}
\label{info_descriptors:sec_descriptors}

\begin{figure*}[tbp]
\begin{center}
\includegraphics[width=\linewidth,clip]{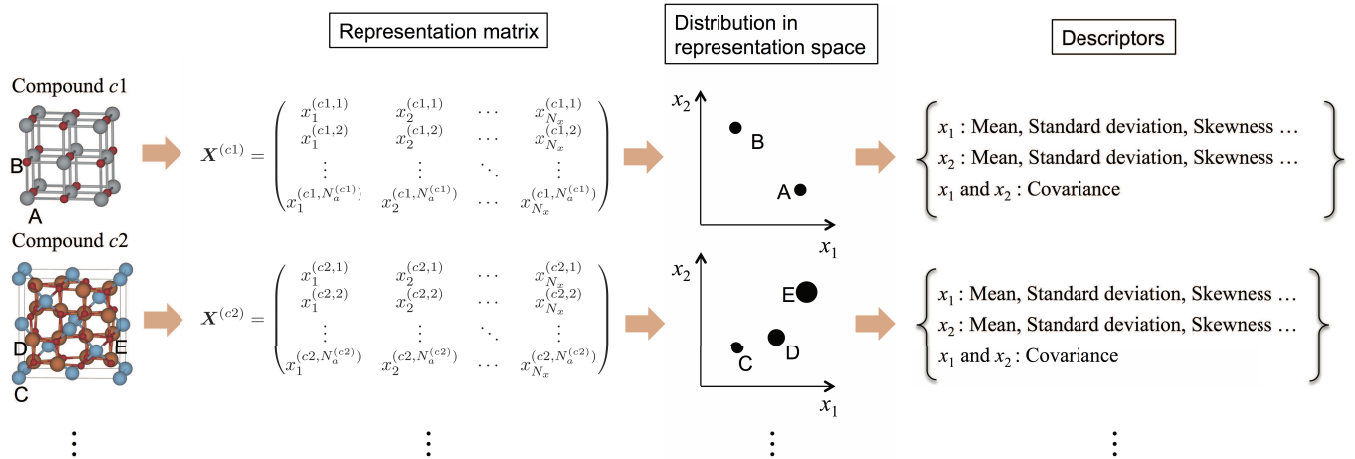} 
\caption{
Schematic illustration of how to generate compound descriptors.
First, each atom in a compound is characterized by $N_x$ representations.
The collection of atoms in the compounds is written as a representation matrix $\bm{X}$.
Then the representation matrix is regarded as the data distribution in an $N_x$-dimensional space.
To transform the distribution into descriptors, representative quantities are introduced to characterize the data distribution such as its mean, standard deviation, skewness, kurtosis and covariance.
}
\label{info_descriptors:schematic_descriptors}
\end{center}
\end{figure*}

\subsection{Representation of compounds}
Here we describe a compound using descriptors $\bm{d}$ that can be derived from only simple elemental and structural representations.
Figure \ref{info_descriptors:schematic_descriptors} schematically illustrates the procedure to generate such descriptors for compounds.
We first consider a compound as a collection of atoms that are described by element types and neighbor environments determined by the other atoms.
Supposing they are represented by $N_{x(\rm ele)}$ elemental representations and $N_{x(\rm st)}$ structural representations, each atom is described by $N_{x} = N_{x(\rm ele)} + N_{x(\rm st)}$ representations.
Therefore, compound $\xi$ is expressed by a collection of the atomic representations as an $(N_a^{(\xi)} \times N_{x})$-dimensional matrix, where $N_a^{(\xi)}$ is the number of atoms included in the unit cell of compound $\xi$.
The representation matrix for compound $\xi$, $\bm{X}^{(\xi)}$, is written as
\begin{equation}
\bm{X}^{(\xi)} = 
\begin{pmatrix}
x_1^{(\xi,1)} & x_2^{(\xi,1)} & \cdots & x_{N_x}^{(\xi,1)} \\
x_1^{(\xi,2)} & x_2^{(\xi,2)} & \cdots & x_{N_x}^{(\xi,2)} \\
\vdots & \vdots & \ddots & \vdots \\
x_1^{(\xi,N_a^{(\xi)})} & x_2^{(\xi, N_a^{(\xi)})} & \cdots & x_{N_x}^{(\xi,N_a^{(\xi)})} \\
\end{pmatrix}
,
\end{equation}
where $x_n^{(\xi,i)}$ denotes the $n$th representation of atom $i$ in compound $\xi$.

Since the representation matrix is only a representation for the unit cell of compound $\xi$, we need a procedure for transforming the representation matrix into a set of descriptors to compare different compounds. 
An approach to the transformation is to regard the representation matrix as the distribution of data points in an $N_x$-dimensional space, as shown in Fig. \ref{info_descriptors:schematic_descriptors}.
To compare the distributions themselves, we then introduce representative quantities to characterize the distribution as descriptors $\bm{d}$, such as the mean, standard deviation, skewness, kurtosis and covariance of the distribution.
The inclusion of the covariance enables the interaction between the element type and crystal structure to be considered.

In previous machine-learning predictions, a popular approach was to use the composition average of a representation as a descriptor, expressed as
\begin{equation}
d_n^{(\xi)} = \frac{1}{N_a^{(\xi)}} \sum_{i=1}^{N_a^{(\xi)}} x_n^{(\xi,i)}.
\end{equation}
In the case of constructing linearized machine-learning interatomic potentials, the average of structural representations is also commonly used because the internal energy is given as the sum of the atomic contributions to the total internal energy\cite{PhysRevB.90.024101}.
The above examples are regarded as simplifications of our approach. 

The performance of this procedure is dependent on the set of elemental representations, the set of structural representations and the representative quantity used to characterize the distribution of the elemental and structural representations.
A universal or complete set of representations that is able to derive a good prediction for all physical properties is desired, while it is expected to be almost impossible to find such a universal set of representations. 
Therefore, we believe that a good approach is to generate as many elemental and structural representations as possible and then select a useful set of representations.

\subsection{Atomic representations}
Our set of elemental representations
\cite{CRC-Handbook-Chemistry-Physics,zunger1981pseudopotential}
is composed of the
(1) atomic number, 
(2) atomic mass, 
(3) period and (4) group in the periodic table,
(5) first ionization energy, 
(6) second ionization energy, 
(7) electron affinity,
(8) Pauling electronegativity, 
(9) Allen electronegativity, 
(10) van der Waals radius, 
(11) covalent radius, 
(12) atomic radius, 
(13) pseudopotential radius for the $s$ orbital, 
(14) pseudopotential radius for the $p$ orbital, 
(15) melting point,
(16) boiling point,
(17) density, 
(18) molar volume, 
(19) heat of fusion,
(20) heat of vaporization,
(21) thermal conductivity and
(22) specific heat.
These representations can be classified into the intrinsic quantities of elements (1)-(7), the heuristic quantities of elements (8)-(14) and the physical properties of elemental substances (15)-(22).

We also introduce four types of structural representations $x_n^{(i)}$, i.e., histogram representations of the partial radial distribution function (PRDF), the generalized radial distribution function (GRDF), the BOP and the angular Fourier series (AFS).
The PRDF is a well-established representation for a wide variety of structures.
Although PRDF has also been used in the context of machine-learning prediction\cite{PhysRevB.89.205118}, it is difficult to apply it directly to a dataset composed of a wide range of compounds in the same way.
Therefore, we apply a histogram representation of the PRDF with a given bin width and cutoff radius to the procedure in this study.
The number of counts for each bin is used as a structural representation.

The GPRF is a pairwise representation similarly to the PRDF, expressed as
\begin{equation}
{\rm GRDF}_n^{(i)} = \sum_j g_n(r_{ij}),
\end{equation}
where $g_n(r_{ij})$ denotes a pairwise function for distance $r_{ij}$ between atom $i$ and its neighbor atom $j$.
For example, a Gaussian pairwise function is given by
\begin{equation}
g_n (r_{ij}) = \exp \left[-a_n(r_{ij} - b_n)^2 \right], 
\end{equation}
where $a_n$ and $b_n$ are the $n$th given parameters.
Here, we employ Gaussian, trigonometric and Bessel pairwise functions as pairwise functions $g_n$.

The GRDF can be regarded as a generalization of the PRDF because the PRDF histogram is obtained by using rectangular functions as pairwise functions $g_n$.
The GRDF has been used not only as a potential function and/or function describing local environment in pairwise interatomic potentials such as Lennard--Jones and embedded atom method (EAM) potentials\cite{lennard1924determination,PhysRevB.29.6443,carlsson1990beyond}, but also as descriptors of machine-learning interatomic potentials\cite{behler2007generalized,PhysRevB.90.024101}.

The BOP is also a well-known representation for local structures in liquid crystal and glass states\cite{PhysRevB.28.784}.
The rotationally invariant BOP $Q_l$ for atomic neighborhoods is expressed by
\begin{equation}
Q_l = \left[ \frac{4\pi}{2l+1} \sum_{m=-l}^{l} |Q_{lm}|^2 \right]^{1/2},
\end{equation}
where $Q_{lm}$ corresponds to the average spherical harmonics for neighbors of atom $i$.
The third-order invariant BOP $W_l$ for atomic neighborhoods is expressed by
\begin{equation}
W_l = \sum^{l}_{m_1, m_2, m_3 = -l}
\begin{pmatrix}
l & l & l \\
m_1 & m_2 & m_3 \\
\end{pmatrix}
Q_{lm_1} Q_{lm_2} Q_{lm_3},
\end{equation}
where the coefficient written by the parentheses is the Wigner 3$j$ symbol, satisfying $m_1+m_2+m_3=0$.
A set of both $Q_l$ and $W_l$ up to a given maximum $l$ is used as structural representations.

The AFS is the most general among the four formulations, which is able to include both the radial and angular dependences of an atomic distribution\cite{bartok2013representing}.
The AFS is given by 
\begin{equation}
{\rm AFS}_{n,n'}^{(i)} = \sum_{j,k} g_n (r_{ij}) g_{n'}(r_{ik}) \cos \theta_{ijk}, 
\end{equation}
where $\theta_{ijk}$ denotes the bond angle between three atoms.
AFS also corresponds to a rotationally-invariant representation simply derived from spherical harmonics modified by radial functions\cite{bartok2013representing}.

%
\section{Results and discussion}
\label{info_descriptors:sec_results}
\subsection{Cohesive energy}

\begin{table}[tbp]
\caption{
Prediction errors of KRR models for the cohesive energy. 
The first row shows the representative quantities of the distribution of atomic representations, where SD stands for the standard deviation.
The first column shows the atomic representations included in the models.
Elemental representations are included in all models.
Values in brackets in the first column are numbers of representations.
Values in brackets in the third column are prediction errors for models with the covariances of atomic representations.
The bottom three lines show the prediction errors of models with structural representations computed using a normalized unrelaxed structure.
(Unit: eV/atom)
}
\label{info_descriptors:rmse_coh}
\begin{ruledtabular}
\begin{tabular}{lcc}
& Mean & Mean $+$ SD \\
\hline
No structural representation & 0.249 & 0.244 (0.231) \\
\hline
Optimized structure & & \\
PRDF (10) & 0.189 & 0.153 (0.110) \\
PRDF (20) & 0.175 & 0.155 (0.106) \\
PRDF (40) & 0.166 & 0.152 (0.125) \\
\hline
Optimized structure & & \\
GRDF (10, trigonometric) & 0.158 & 0.104 (0.050) \\
GRDF (20, trigonometric) & 0.158 & 0.094 (0.045) \\
GRDF (40, trigonometric) & 0.149 & 0.093 (0.053) \\
GRDF (10, Gaussian) & 0.170 & 0.108 (0.056) \\
GRDF (20, Gaussian) & 0.166 & 0.101 (0.058) \\
GRDF (40, Gaussian) & 0.157 & 0.100 (0.051) \\
GRDF (80, Gaussian) & 0.156 & 0.099 (0.061) \\
GRDF (10, Bessel) & 0.172 & 0.106 (0.055) \\
GRDF (20, Bessel) & 0.169 & 0.104 (0.055) \\
\hline
Optimized structure & & \\
BOP (20) & 0.156 & 0.129 (0.064) \\ 
BOP (20) & \multirow{2}{*}{0.108} & \multirow{2}{*}{0.077 (0.041)} \\
+ GRDF (20, trigonometric) & & \\
AFS (10, trigonometric) & 0.139 & 0.102 (0.079) \\
AFS (20, trigonometric) & 0.146 & 0.102 (0.103) \\
\hline
Normalized unrelaxed structure & & \\
PRDF (20) & 0.166 & 0.162 (0.072) \\
PRDF (40) & 0.166 & 0.164 (0.071) \\
GRDF (20 trigonometric) & 0.169 & 0.164 (0.074) \\
\end{tabular}
\end{ruledtabular}
\end{table}

To begin with, the performance of descriptors is examined by developing KRR prediction models for the DFT cohesive energy.
First, we adopt descriptor sets derived only from elemental representations, which are expected to be more important than structural representations in the prediction of the cohesive energy.
Since the elemental representations are not complete for some of the elements in the dataset, we consider only the elemental representations that are complete for all elements.
We estimate the root-mean-square error (RMSE) for test data composed of a randomly selected 10\% of the data.
The random selection of test data is repeated 20 times, and the average RMSE is regarded as the prediction error.

\begin{figure}[tbp]
\begin{center}
\includegraphics[width=0.95\linewidth,clip]{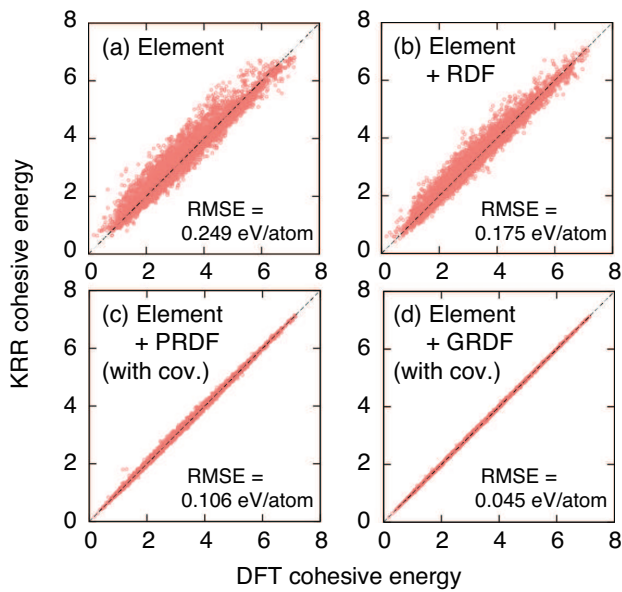} 
\caption{
Comparison of cohesive energy calculated by DFT calculation and that calculated by the KRR prediction model.
Only one test dataset is shown.
Descriptor sets are composed of 
(a) the means of elemental representations, 
(b) the means of elemental and PRDF representations,
(c) the means, SDs and covariances of elemental and PRDF representations
and (d) the means, SDs and covariances of elemental and 20 trigonometric GRDF representations.
The mean of the PRDF corresponds to the RDF.
Structure representations are computed from the optimized structure for each compound.
}
\label{info_descriptors:cohesive_comparison}
\end{center}
\end{figure}

Table \ref{info_descriptors:rmse_coh} summarizes the prediction errors of KRR models for the cohesive energy.
The simplest option is to use only the mean of each elemental representation as a descriptor.
The prediction error in this case is 0.249 eV/atom. 
When considering the means, standard deviations and covariances of elemental representations, the prediction model has a prediction error of 0.231 eV/atom.
Figure \ref{info_descriptors:cohesive_comparison} (a) shows a comparison of the cohesive energy calculated by DFT calculation and by the best KRR model, where only the test data in one of the 20 trials are shown.
As can be seen in Fig. \ref{info_descriptors:cohesive_comparison} (a), many data largely deviate from the diagonal line representing equal DFT and KRR energies.
We also found that the skewness and kurtosis are not important descriptors for the prediction.

Since we consider several crystal structures for each chemical composition, all the data have an intrinsic error originating from the absence of structural representations.
The intrinsic standard deviation averaged over the chemical compositions $\sigma_{\rm int}$ can be estimated as
\begin{equation}
\sigma_{\rm int} = \frac{1}{N_{\rm comp}} \sum_i {\sqrt {\frac{1}{N_i^{(\rm st)}} \sum_s \left(E_{i,s} - \langle E_i \rangle \right)^2} },
\end{equation}
where $N_{\rm comp}$ and $N_i^{(\rm st)}$ denote the number of chemical compositions and the number of prototype structures for chemical composition $i$, respectively.
$E_{i,s}$ and $\langle E_i \rangle$ are the cohesive energy for chemical composition $i$ with prototype structure $s$ and the average cohesive energy for chemical composition $i$, respectively.
The intrinsic standard deviation is estimated to be 0.211 eV/atom, which is close to the prediction error for all models, indicating that our set of elemental representations is nearly complete for the prediction of the cohesive energy.

Next, we introduce descriptors related to structural representations.
They can be computed from both the crystal structure optimized by the DFT calculation and the initial prototype structures.
The former is only useful for machine-learning prediction when an observation is expensive.
Since the optimized structure calculation requires the same computational cost as the cohesive energy calculation, the benefit of using machine learning is lost when using the optimized structure.
Only to examine the limitation of the procedure and the atomic representations introduced in this study, structural representations are first computed from the optimized crystal structure.
KRR models are constructed using many descriptor sets composed of elemental and structural representations.
The cutoff radius is set to 6 \AA\ for the PRDF, GRDF and AFS, and the cutoff radius is set to 1.2 times the nearest-neighbor distance for the BOP, which is a widely used definition of the nearest neighbor.

First, three sets of PRDF histogram representations are applied.
The number of histogram representations for each set is controlled only by the bin width.
Figure \ref{info_descriptors:cohesive_comparison} shows a comparison of the DFT and KRR cohesive energies, where the KRR models are constructed by (b) a set of the means of the elemental and PRDF representations and (c) a set of the means, standard deviations and covariances of the elemental and PRDF representations.
When only considering the means of the elemental and PRDF representations, the lowest prediction error is as large as 0.166 eV/atom, as shown in Table \ref{info_descriptors:rmse_coh}.
This means that the simple use of the PRDF does not enable a good model for the cohesive energy to be developed.
However, by including the covariances of the elemental and PRDF representations, a much better prediction model is obtained and the prediction error significantly decreases to 0.106 eV/atom.

Table \ref{info_descriptors:rmse_coh} also shows the prediction error of KRR models with descriptors obtained from the elemental and GRDF representations.
Considering only the means of the GRDFs, we obtain prediction models with errors of 0.149$-$0.172 eV/atom, which are very close to those of the prediction models considering the means of the PRDFs.
Similarly to in the case of the PRDF, the prediction model is improved by considering the SDs and covariances of the elemental and structural representations.
The best model shows a prediction error of 0.045 eV/atom, which is about half of that of the best PRDF model.
This is also approximately equal to the ``chemical accuracy" of 43 meV/atom (1 kcal/mol).
Figure \ref{info_descriptors:cohesive_comparison} (d) shows a comparison of DFT and KRR cohesive energies, where a set of the means, SDs and covariances of the elemental and trigonometric GRDF representations is adopted. 
As can be seen in Fig. \ref{info_descriptors:cohesive_comparison} (d), most of the data are located near the diagonal line.
Table \ref{info_descriptors:rmse_coh} also shows the prediction error of KRR models with descriptor sets including the angular-dependent structural representations of the BOPs and AFSs.
We obtain the best prediction model with a prediction error of 0.041 eV/atom by considering the means, SDs and covariances of the elemental, 20 trigonometric GRDF and 20 BOP representations.

\begin{figure}[tbp]
\begin{center}
\includegraphics[width=0.85\linewidth,clip]{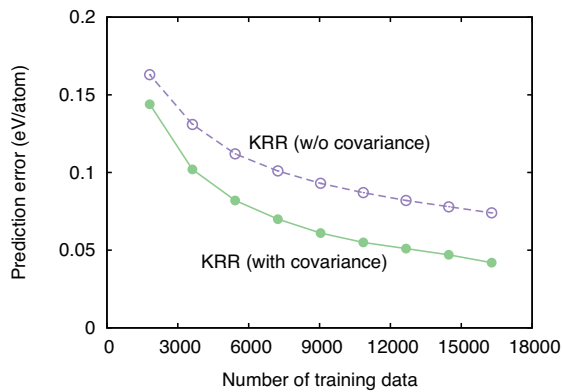} 
\caption{
Dependence of the prediction error on the number of training data.
A set of 20 trigonometric GRDFs and 20 BOPs is used as the structural representations.
Open circles and closed circles show the prediction error of the KRR model without and with consideration of the covariances of the elemental and structural representations, respectively.
The numbers of descriptors are 122 and $1952$ in the former and latter cases, respectively. 
}
\label{info_descriptors:n_obs_ecoh}
\end{center}
\end{figure}

We also examine the dependence of the prediction error on the number of training data, as shown in Fig. \ref{info_descriptors:n_obs_ecoh}.
A set of 20 trigonometric GRDFs and 20 BOPs is used as structural representations, which derives the best model of the cohesive energy.
The prediction error gradually decreases with increasing number of training data.
Note that the prediction error of the KRR model with the covariance is close to that of the KRR model without the covariance when the number of training data is small.
In other words, a better prediction is expected by including covariances only when a large training dataset is available. 
This may be ascribed to the fact that the inclusion of the covariances leads to a significant increase in the number of descriptors.
The same behavior can be found in the following sections on the predictions of the LTC and melting temperature using other small datasets.

We used the optimized structure to calculate structural representations. 
To make a practical prediction model for the cohesive energy, however, it is necessary to use structures that can be obtained without performing the DFT calculation. 
Therefore, we apply normalized prototype structures obtained without performing the DFT calculation.
The prototype structures are isotropically normalized so that the volume per atom becomes 1 \AA$^3$.
Table \ref{info_descriptors:rmse_coh} shows the prediction errors of KRR models with descriptor sets based on the normalized prototype structural representations.
Contrary to the use of the optimized structure, the prediction models with trigonometric GRDFs and PRDFs have almost the same prediction error.
The prediction error of the model with the means of the elemental and structural representations is 0.166 eV/atom, which is almost the same as those of the prediction models with the optimized structural representations.
At the same time, it is important to consider the covariances of the elemental and structural representations, which improve the prediction error to about 0.07 eV/atom.

\begin{table}[tbp]
\caption{
Prediction errors of KRR models for LTC and melting temperature.
}
\label{info_descriptors:rmse_prop}
\begin{ruledtabular}
\begin{tabular}{lcc}
& Mean & Mean + SD \\
\hline
LTC & & \\
No structural representation & 0.173 & 0.142 (0.130) \\
GRDF (20) & 0.179 & 0.108 (0.137) \\
BOP (20) & 0.128 & 0.096 (0.155) \\
GRDF (20) + BOP (20) & 0.156 & 0.102 (0.149) \\ 
\hline
Melting temperature & & \\
No structural representation & 278 & 273 (236) \\
GRDF (20) & 302 & 277 (301) \\
BOP (20) & 264 & 238 (286) \\
GRDF (20) + BOP (20) & 293 & 278 (307) \\ 
\end{tabular}
\end{ruledtabular}
\end{table}

\subsection{LTC}
The amount of training data is limited for most physical properties of interest.
Therefore, we believe that it is very important to examine the performance of descriptors using a small dataset when employed in physics and materials science.
Here both KRR and Bayesian optimization are performed to examine the performance of descriptors using log-scaled LTC data.
Since the computational cost of obtaining the optimized DFT structure is much smaller than that for obtaining the LTC, we use the optimized DFT structure to compute structural representations.
The prediction error is estimated as the RMSE for test data, which is composed of a randomly selected 10\% of the data.
The random selection of the test data is repeated 200 times, and then the average RMSE is regarded as the prediction error.

Table \ref{info_descriptors:rmse_prop} shows the prediction errors of KRR models for the LTC data.
As well as achieving cohesive energy prediction, the structural representations improve the prediction model.
On the other hand, the inclusion of the covariances reduces the accuracy of the prediction model due to the small training dataset, as observed in the previous section.
The best model is composed of the means and SDs of the elemental and BOP representations, having a prediction error of $0.096$ for the log-scaled LTC.

\begingroup
\squeezetable
\begin{table}[tbp]
\caption{
Performance of Bayesian optimization using the LTC data.
Average numbers of samples required to find PbClBr, CuCl and LiI are shown, which have the lowest, 11th-lowest and 12th-lowest LTCs among the 110 compounds, respectively.
The means and SDs of the representations are considered in all models.
$+$ and $-$ in the first three columns show the representations and covariances included and not included in the prediction models, respectively.
}
\label{info_descriptors:bo_ltc_table}
\begin{ruledtabular}
\begin{tabular}{cccccccccccccccc}
\multirow{2}{*}{GRDF} & \multirow{2}{*}{BOP} & \multirow{2}{*}{Cov.} & \multirow{2}{*}{RMSE} & \multicolumn{2}{c}{PbClBr} & \multicolumn{2}{c}{CuCl} & \multicolumn{2}{c}{LiI} & \multicolumn{2}{c}{Average} \\
& & & & PI & EI & PI & EI & PI & EI & PI & EI \\
\hline
$-$ & $-$ & $-$ & 0.142 & 13.9 &  12.3 &  54.4 &  47.5  & 17.5 &  18.9 &  28.6 &  26.2 \\
$+$ & $-$ & $-$ & 0.108 & 7.6  &  7.9  &  40.5 &  41.0  & 48.6 &  49.7 &  32.2 &  32.9 \\
$-$ & $+$ & $-$ & 0.096 & \textbf{5.0}  &  5.2  &  15.1 &  15.7  & \textbf{9.1}  &  9.4  &  \textbf{9.7}  &  10.1 \\
$+$ & $+$ & $-$ & 0.102 & 5.0  &  5.1  &  22.4 &  22.3  & 28.4 &  27.0 &  18.6 &  18.1 \\
$-$ & $-$ & $+$ & 0.130 & 35.0 &  32.3 &  11.8 &  11.4  & 30.1 &  33.1 &  25.6 &  25.6 \\
$+$ & $-$ & $+$ & 0.137 & 8.8  &  8.7  &  31.8 &  31.7  & 84.5 &  83.9 &  41.7 &  41.4 \\
$-$ & $+$ & $+$ & 0.155 & 13.7 &  14.2 &  \textbf{8.9}  &  \textbf{9.0}   & 43.4 &  44.2 &  22.0 &  22.5 \\
$+$ & $+$ & $+$ & 0.149 & 9.0  &  9.0  &  13.9 &  14.1  & 63.1 &  64.0 &  28.7 &  29.0 \\
\hline
\multicolumn{4}{c}{Random}  & \multicolumn{2}{c}{50} & \multicolumn{2}{c}{55} & \multicolumn{2}{c}{55} & \multicolumn{2}{c}{$-$}
\end{tabular}
\end{ruledtabular}
\end{table}
\endgroup

\begin{figure}[tbp]
\begin{center}
\includegraphics[width=\linewidth,clip]{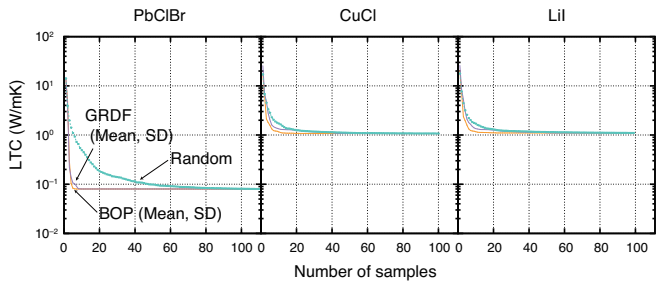} 
\caption{
Behavior of Bayesian optimization for the LTC data in finding PbClBr, CuCl and LiI.
The best LTC is shown along with the number of samples (iterations) during the Bayesian optimization and random search.
}
\label{info_descriptors:bo_ltc_figure}
\end{center}
\end{figure}

Next, the performance of Bayesian optimization is examined.
Both Bayesian optimization and random searches are repeated 200 times and the average number of samples required to find the compound with the lowest LTC, i.e., PbClBr, is examined.
Table \ref{info_descriptors:bo_ltc_table} shows the performance of Bayesian optimization using the PI and EI algorithms.
The means and SDs of the representations are considered in all models.
Figure \ref{info_descriptors:bo_ltc_figure} shows the behavior of the lowest LTC during Bayesian optimization in comparison with that in the random search.
When using the GP model with the BOP, the average number of samples required for the optimization, $N_{\rm ave}$, is 5.0, which is ten times smaller than that of the random search, $N_{\rm ave}=50$.
PbClBr is discovered much more efficiently by Bayesian optimization than by the random search.

To evaluate the performance for finding a wide variety of low-LTC compounds, we prepare two datasets after intentionally removing some low-LTC compounds.
In these datasets, CuCl and LiI, respectively showing 11th-lowest and 12th-lowest LTCs, are the solutions of the optimizations.
Table \ref{info_descriptors:bo_ltc_table} also shows the result of Bayesian optimization using the datasets.
When using the GP model with BOPs, the average numbers of observations required to find CuCl and LiI are $N_{\rm ave}=15.1$ and $9.1$, respectively, which are much smaller that those of the random search.
On the other hand, when using the GP model with GRDFs, the average numbers of observations required to find CuCl and LiI are $N_{\rm ave}=40.5$ and $48.6$, respectively.
The delay of the optimization may originate from the fact that both CuCl and LiI may be outliers in the model with GRDFs, although the model with GRDFs has a similar RMSE to that of the model with BOPs.
These results indicate that we need to optimize a set of descriptors by examining the performance of Bayesian optimization for a wide range of compounds to find such outlier compounds.

\subsection{Melting temperature}

\begingroup
\squeezetable
\begin{table}[tbp]
\caption{
Performance of Bayesian optimization using the melting temperature data.
Average numbers of observations required to find AlN, SiC and MgO are shown, which exhibit the lowest, second-lowest and third-highest melting temperatures among the 248 compounds, respectively.
The means and SDs of the representations are considered in all models.
$+$ and $-$ in the first three columns show the representations and covariances included and not included in the prediction models, respectively.
}
\label{info_descriptors:bo_mp_table}
\begin{ruledtabular}
\begin{tabular}{cccccccccccccccc}
\multirow{2}{*}{GRDF} & \multirow{2}{*}{BOP} & \multirow{2}{*}{Cov.} & RMSE & \multicolumn{2}{c}{AlN} & \multicolumn{2}{c}{SiC} & \multicolumn{2}{c}{MgO} & \multicolumn{2}{c}{Average} \\
& & &(K) & PI & EI & PI & EI & PI & EI & PI & EI \\
\hline
$-$ & $-$ & $-$ & 273 & 38.9 & 39.1 & 26.0 & 26.6 & 29.4 & 27.7 & 31.4 & 31.1 \\
$+$ & $-$ & $-$ & 277 & 24.5 & 25.4 & 30.9 & 36.8 & 22.8 & 25.8 & 26.1 & 29.3 \\
$-$ & $+$ & $-$ & 238 & 22.5 & 22.5 & 28.2 & 32.0 & 20.9 & 21.1 & 23.9 & 25.2 \\
$+$ & $+$ & $-$ & 278 & 30.4 & 34.3 & 35.4 & 42.6 & 30.7 & 34.0 & 32.2 & 37.0 \\
$-$ & $-$ & $+$ & 236 & 27.6 & 27.7 & 52.1 & 53.2 & 28.8 & 28.4 & 36.2 & 36.4 \\
$+$ & $-$ & $+$ & 301 & 35.5 & 41.8 & 71.8 & 73.5 & 35.0 & 43.8 & 47.4 & 53.0 \\
$-$ & $+$ & $+$ & 286 & 38.4 & 48.0 & 68.5 & 71.6 & 35.2 & 39.2 & 47.4 & 52.9 \\
$+$ & $+$ & $+$ & 307 & 50.5 & 68.1 & 69.8 & 84.3 & 41.4 & 48.4 & 53.9 & 66.9 \\
\hline
\multicolumn{4}{c}{Random}  & \multicolumn{2}{c}{125} & \multicolumn{2}{c}{125} & \multicolumn{2}{c}{125} & \multicolumn{2}{c}{$-$}
\end{tabular}
\end{ruledtabular}
\end{table}
\endgroup

\begin{figure}[tbp]
\begin{center}
\includegraphics[width=\linewidth,clip]{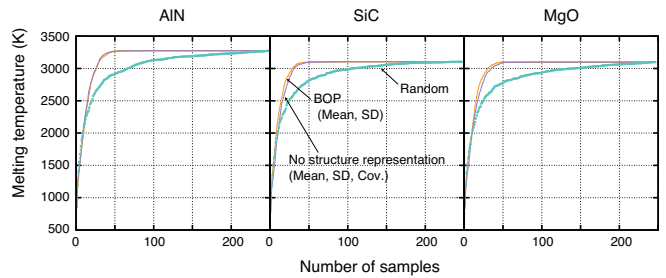} 
\caption{
Behavior of Bayesian optimization for the melting temperature data in finding AlN, SiC and MgO.
}
\label{info_descriptors:bo_mp_figure}
\end{center}
\end{figure}

The other example of a small dataset is the experimental melting temperature of solids.
KRR and Bayesian optimization are performed as well as the LTC prediction.
We use the optimized DFT structure to compute structural representations.
Similarly to in the LTC prediction, we use test data composed of a randomly selected 10\% of the data to estimate the average prediction error over 200 random selections of the test data.
Table \ref{info_descriptors:rmse_prop} shows the prediction errors of KRR models for the melting temperature dataset.
In contrast to the cohesive energy and LTC predictions, only the BOP improves the prediction model, while the GRDF reduces its accuracy.
The inclusion of the covariances improves the prediction model only when considering elemental representations.
This may be ascribed to the small number of training data, similarly to in the case of the LTC.
The best model is composed of the means, SDs and covariances of elemental information and has a prediction error of $236$ K.

Finally, the performance of Bayesian optimization to find a target compound is examined.
The number of observations required to find a target compound averaged over 200 trials is evaluated for both Bayesian optimization and a random search.
Table \ref{info_descriptors:bo_mp_table} shows the number of observations required to find a target compound by Bayesian optimization compared with that for a random search. 
To evaluate the performance in finding a wide variety of high-melting-temperature compounds, we prepare two datasets after intentionally removing some high-melting-temperature compounds.
In this case, SiC and MgO with the second-highest and third-highest melting temperatures, respectively, are the solutions of the optimization.

Figure \ref{info_descriptors:bo_mp_figure} shows the behavior of the highest melting temperature among the samples during Bayesian optimization to find AlN, SiC and MgO in comparison with the random search.
When using the best RMSE model, composed of the mean, SD and covariances of the elemental representations, the average numbers of samples required to find AlN, SiC and MgO are $N_{\rm ave}=27.6$, $52.1$ and $28.8$, respectively.
When using the second-best RMSE model composed of the mean and SD of the elemental and BOP representations, with an RMSE close to that of the best model, the average numbers of samples required to find AlN, SiC and MgO are $N_{\rm ave}=22.5$, $28.2$ and $20.9$, respectively, which are about six times smaller than the numbers of samples required to find the targets in the random search.

\section{Conclusion}
\label{info_descriptors:sec_conclusion}
In this study, we have demonstrated an approach to generate a systematic set of compound descriptors from simple atomic representations.
It was applied to three datasets for the cohesive energy, LTC and experimental melting temperature.
We examined the performance of the sets of descriptors in terms of the accuracy of the kernel ridge models and the performance of Bayesian optimization.
For the cohesive energy dataset, we obtained the best prediction model with a prediction error of 0.041 eV/atom, which is approximately equal to the ``chemical accuracy" of 1 kcal/mol (0.043 eV/atom).
Also in the predictions of the LTC and melting temperature, the present method exhibits good performances for both the kernel ridge models and Bayesian optimization.
Although we focused on crystalline compounds, the structure descriptors used in the present study can be easily applied to noncrystalline or molecular compounds. 
The present method should therefore be useful for searching for compounds with many different chemical properties and applications from a wide range of chemical and structural spaces without performing exhaustive DFT calculations.

\begin{acknowledgments}
This work was supported by PRESTO, JST and a Grant-in-Aid for Scientific Research on Innovative Areas ``Nano Informatics" (Grant No. 25106005) from the Japan Society for the Promotion of Science (JSPS).
AS, HH and IT were also supported by the ``Materials Research by Information Integration" Initiative (MI$^2$I) from Japan Science and Technology Agency.
IT was also supported by a Grant-in-Aid for Scientific Research (A) from JSPS.
\end{acknowledgments}

\bibliography{informatics_descriptors}

\end{document}